\documentclass[12pt]{JHEP3}
\usepackage{graphicx}
\usepackage{dcolumn}
\usepackage{bm}
\usepackage{amsmath}
\usepackage{amsfonts}
\usepackage{citesort}

\def\drawbox#1#2{\hrule height#2pt
        \hbox{\vrule width#2pt height#1pt \kern#1pt
              \vrule width#2pt}
              \hrule height#2pt}

\def\Asym#1#2{\vcenter{\vbox{\drawbox{#1}{#2}
              \kern-#2pt       
              \drawbox{#1}{#2}}}}

\newcommand{\beq}{\begin{eqnarray}}
\newcommand{\eeq}{\end{eqnarray}}

\usepackage{verbatim}

\title{Very Large Primordial Non-Gaussianity from multi-field:
Application to Massless Preheating}

\author{Asko Jokinen,\\ NORDITA, Blegdamsvej-17,
Copenhagen-2100, Denmark.}  \author{Anupam Mazumdar,\\ NORDITA,
Blegdamsvej-17, Copenhagen-2100, Denmark.}

\abstract{In this paper we derive a generic expression, which is valid
for scales larger than Hubble radius and contains only the local
terms, for the second order curvature perturbations for more than one
field, provided the expansion is sourced by the energy density of a
single field. As an application, motivated by our previous
paper~\cite{asko1}, we apply our formalism to two fields during
preheating, where the inflaton oscillations are sourced by
$\lambda\varphi^4$ potential which is governing the expansion of the
Universe. A second field $\sigma$, coupled to the inflaton through
$g^2\varphi^2\sigma^2$, is excited from the vacuum fluctuations. The
excited modes of $\sigma$ amplify the super-Hubble isocurvature
perturbations, which seed the second order curvature perturbations to
give rise to a significantly large non-Gaussianity. Our results show
that within $3$ inflaton oscillations for a range of parameters, $1<
g^2/\lambda < 3$, the non-Gaussianity parameter becomes: $f_{NL}\geq
{\cal O}(1000)$, which is already ruled out by the current WMAP
observation.}

\preprint{NORDITA-2005-79} 

\begin{document}

\section{Introduction}

Inflation stretches the fluctuations outside the Hubble radius which
later on enter inside the Hubble radius to seed the structure
formation~\cite{Liddle-Lyth}.  A single field inflationary model
generates Gaussian fluctuations which is in an excellent agreement
with the current observations~\cite{WMAP}. These Gaussian fluctuations
are due to the adiabatic spectrum of the curvature
perturbations~\cite{Mukhanov}, for a review see~\cite{Brandenberger},
which in a single field case is conserved after the wavelength of a
perturbed mode becomes larger than the Hubble
radius~\footnote{Perturbations whose wavelength is larger than the
Hubble radius are often abused in the literature as ``super-horizon
perturbations'', instead they should be called as ``super-Hubble
perturbations.  Note that due to inflation the perturbations are
always well inside the causal horizon. We particularly thank Robert
Brandenberger for stressing this point time and again.}.  For multiple
fields there is an iso-curvature component which can source the
curvature perturbations, i.e., as a result the curvature perturbations
are no longer conserved on large scales.

Although negligible but a finite calculable departure from Gaussianity
can be obtained within a single field inflationary model, with a self
interacting inflaton potential.  However the measure for
non-Gaussianity, defined as $f_{NL}$ (see the precise definition
below), is bounded by the slow roll parameters~\cite{Acquaviva,Maldacena}, i.e.,
$f_{NL}\sim |\eta-3\epsilon|$, where the slow roll parameters,
$\epsilon,~\eta \ll 1$. This trend remains true even for multi field case,
see for example ~\cite{Enqvist,Lidsey,Lyth1}. Unless the slow roll
conditions are violated during inflation, it is hard to imagine a
reasonably large non-Gaussianity can be generated, which would really
challenge the common lore that slow roll inflation produces mainly
Gaussian fluctuations. For a review on non-gaussianity, see \cite{Bartolo:2004if}.

The only exceptional moment is right at the time when inflation ends, when the slow roll parameters
become large, i.e., $\epsilon,~\eta \sim {\cal O}(1)$, nevertheless this still
leaves $f_{NL}\sim {\cal O}(1)$ \cite{Lyth2}. Large $f_{NL}\gg 1$, however, can be obtained
during preheating after inflation as we illustrated in our previous
paper~\cite{asko1}~\footnote{See also other examples considered in
Refs.~\cite{asko2,asko3} and \cite{Finelli:2001db}.}.

Particularly a scalar preheating~\cite{some} is a phase where after
inflation the coherent oscillations of the inflaton, $\varphi$,
dominates the energy density of the Universe and another scalar field,
$\sigma$, coupled to the inflaton through, $g^2\varphi^2\sigma^2$, is
excited from the vacuum fluctuations. In this respect preheating is a
non-perturbative phenomena.  Although preheating does not lead to a
complete thermalization~\cite{averdi,averdi1}, nevertheless, in many
models preheating can be just a possibility before the perturbative
decay of the inflaton, for example see the last reference
of~\cite{some}, if there exists favorable conditions such as
significantly large coupling.  In reality it is hard to predict
exactly how the inflaton couples to the Standard Model degrees of
freedom, due to lack of a model where the inflaton is not an absolute
gauge singlet~\cite{Enqvist-anu}~\footnote{An exceptional case of
assisted inflation~\cite{assist} with gauge invariant supersymmetric
flat directions~\cite{Enqvist-anu} can possibly address this
crisis~\cite{asko-anu}.}, therefore, it becomes difficult to pin down
thermalization time scales, etc., which can be tested through present
cosmological experiments.  Nevertheless, very recently as we pointed
out that preheating can give rise to a large non-Gaussianity which can
lead to constraining certain parameter regions, such as a coupling
constant~\cite{asko1,asko2,asko3}, we regard this as a window of
opportunity to test the initial stage of preheating.

In order to understand why preheating is the test bed for a large
primordial non-Gaussianity. Let us first note that in general we
require:

\begin{itemize}

\item{Significant interaction which leads to curved potential that
breaks slow roll conditions.}

\item{Some small amplitude isocurvature fluctuations which could be
amplified later on via non-adiabatic evolution.}

\end{itemize}

During preheating it is possible to have both the criteria
fulfilled. The first one can be obtained mainly because during
preheating the slow roll conditions are violated. The second condition
is also satisfied because the $\sigma$ field, which carries
isocurvature fluctuations~\footnote{The fluctuations in $\sigma$ field
will be scale invariant and the amplitude of the fluctuations in
$\sigma$ will be similar to that of the inflaton, i.e.,
$\delta\sigma\sim H_{inf}/2\pi$ (during inflation and after the first
few inflaton oscillations). With these comparatively large initial
amplitude and scale invariant fluctuations it is possible to amplify
the fluctuations in $\delta \sigma_{k}$ on super Hubble scales.} ,
undergoes non-adiabatic evolution.  The quantum modes of $\sigma$
field are excited because of the time dependence on the effective mass
of $\sigma$ field (due to coupling between $\varphi$ and $\sigma$
where $\varphi$ is oscillating) changes sign, therefore violates
adiabatic evolution of the vacuum.

Still a reader might wonder why amplifying $\delta\sigma_{k}$ leads to
a large non-Gaussianity? After all non-Gaussianity is related to the
second order curvature fluctuations. Two specific questions could be:

\begin{itemize}

\item{Can a sub-Hubble process influence the super-Hubble second order
curvature perturbations?}

\item{Can existing super-Hubble modes amplify themselves to influence 
the second order curvature perturbations?}

\end{itemize}

Note that the second order perturbations take into account of the
interactions in leading order of the perturbation theory, namely the
quadratic combinations of the first order perturbations act as sources
for the second order. In momentum space the sources become
convolutions and thus couple different scales. To answer the first
question, we should also keep in mind that inflation makes the {\it
causal horizon} exponentially large compared to the Hubble
radius. Therefore causality allows various modes to mix, i.e.,
super-Hubble modes in second order can be influenced by either
convoluting the first order sub-Hubble modes or the first order
super-Hubble modes. In paper~\cite{asko1} we studied preheating
related to $m^2\varphi^2$-potential for the inflaton where the
resonance affects the super-Hubble modes in second order through
sub-Hubble perturbations, i.e., it relates to the first question.

In this paper we provide a generic expression for the second order
curvature perturbation which are fed by the super-Hubble fluctuations
in the first order. As an example we consider preheating in
$\lambda\varphi^4$-potential for the inflaton where the second order
super-Hubble modes are affected.

Let us now highlight the role of isocurvature fluctuations. It is a
well known fact that the isocurvature fluctuations source the
adiabatic fluctuations and therefore the curvature perturbations.  The
adiabatic perturbations cannot increase by themselves on super-Hubble
scales without iso-curvature perturbations.  This is true in first
order perturbations, in second order case there are additional sources
which feed the second order metric and curvature perturbations, we
shall see them below. Without having isocurvature fluctuations it is
therefore not possible to amplify the first order and subsequently the
second order curvature perturbations. However note that
non-Gaussianity is a ratio between the second order curvature
perturbation with respect to square of the first order curvature
perturbations. If both first and second order curvature perturbations
grow due to large isocurvature fluctuations then the ratio need not be
large always.

Therefore in order to get large non-Gaussianity we need a setup where
only the second order curvature perturbations grow while the first
order curvature perturbations do not grow at all. Can this happen in
reality?

Yes indeed, this can happen when the isocurvature fluctuations do not
seed the first order perturbations at all, but it only seeds the
second order perturbations. We achieved this modest goal in our first
paper,~\cite{asko1}, where the background motion of a $\sigma$ field
were absent, only the coherent oscillations of $\varphi$ field were
present. In which case the perturbations in $\sigma$ decouple from the
first order perturbations, but not in the second order and in higher
order perturbation theory. Therefore exponentially large growth in
$\delta\sigma_{k}$ can feed the second order curvature perturbations
to give rise to a very large non-Gaussianity.

The aim of the present paper is to lay down the formalism for second
order curvature perturbations, where there is a single field
dominating the energy density of the Universe while other multi-fields
can leave their imprints via fluctuations~\footnote{In our previous
papers~\cite{asko1,asko2,asko3} we mainly studied the first order and
second order metric perturbations, which need not be conserved on
large scales. Here our formalism is more robust as we study the
perturbations in terms of conserved quantities,
$\zeta^{(1)},~\zeta^{(2)}$, for the definition see below.  In this
paper we verify our earlier claim that preheating can really boost
large primordial non-Gaussianity.}.  We also show that the second
order curvature perturbation becomes {\it free} from the {\it
non-local contributions} on super-Hubble scales when we neglect the
gradient terms, which is true in a generic background (all the scalar
fields can have non-vanishing VEVs), see the detailed discussion
in~\ref{NLT}.

As an application we will consider a massless preheating, where the
$\sigma$ field does not have a bare mass term and the inflaton
potential is quartic, i.e., $V(\varphi,\sigma)\sim
\varphi^4+g^2\varphi^2\sigma^2$. This potential has certain nice
properties which we will highlight in the course of our discussion,
i.e., the instability band for the $\delta\sigma_k$ mode depends on
the ratio, $g^2/\lambda$, cosmologically interesting solutions can be
obtained for the first instability band, $1< g^2/\lambda < 3$.  In
order to obtain the results analytically, we assume that the
background value of $\sigma $ is vanishing. A non-vanishing case has
to be studied separately with an aid of numerics~\footnote{If $\sigma$
field is also oscillating and contributing to the energy density of
the Universe then again we would have to address the problem
numerically.}.

A large non-Gaussianity also depends on the number of oscillations, if
the number of oscillations grow the non-Gaussianity grows
exponentially at an alarming rate, see the
Plots.~[\ref{lphi4f1},\ref{lphi4f2}]. A simple explanation can be
given as follows, fluctuations of the $\sigma$ field grow as $\sim
e^{2\mu x}$, where $\mu\sim 0.2$ is a physical parameter governing the
production of $\sigma$ particles ( for a detailed discussion, see
below) and $x$ is time in dimensionless units which measures the
period of oscillations, $T\sim 7$, so that the time elapsed is the
number of oscillations, $N\sim 3$, times the period, i.e.,
$x=NT$. With these numbers the growth factor is amazingly large,
\begin{equation}
e^{2\mu x}\sim e^{8.4}\sim 4000\,. 
\end{equation}\label{growth}
Barring accidental cancellations between various source terms or
dampening the initial amplitude for the isocurvature fluctuations this
leads to a large non-Gaussianity.

However we should remind the readers that the phase of preheating
cannot last forever. Besides the metric and curvature perturbations
there is a backreaction on energy momentum components due to
exponentially large particle creation, this inevitably stops
preheating. Depending on the couplings the backreaction can stop
preheating in dozens of oscillations. When the backreaction becomes
important our assumption with a vanishing VEV of $\sigma$ field also
breaks down.

Unless the time evolution of $\sigma$ field really brings down the
non-Gaussianity parameter to the level that is consistent with the
current observations, the model considered in our paper with $1<
g^2/\lambda<3$ is deemed to be ruled out. For other instability bands
the effective mass for the $\sigma $ field becomes heavy, i.e.,
$m_{\sigma,~eff}\geq H_{inf}$, which decreases the amplitude of the
isocurvature fluctuations and leads to a smaller non-Gaussianity
parameter. In this respect non-Gaussianity acts as a discriminatory and
it should be considered as a useful observable to distinguish various
physical situations.

We begin with the definitions, then we describe the first and second
order curvature perturbations. We then discuss the background
evolution where we obtain the final expression for the second order
curvature perturbations for more than one field, provided the
background evolution is governed by one field alone, for an example
this could be the inflaton. We then turn our attention to a massless
preheating and derive an expression for the non-Gaussianity parameter.
In order to be self consistent, we added an appendix, where useful
relationships have been derived explicitly.


\section{Basic equations}
\subsection{Definitions}
The metric tensor, including only the scalar perturbations upto the
second order, is given by~\cite{Malik:2005cy}
\begin{eqnarray}
\label{metric1}
g_{00}&=&-a^2\left(1+2\phi^{(1)}+\phi^{(2)}\right) \,, \\
g_{0i}&=&a^2\left(B^{(1)}+\frac{1}{2}B^{(2)}\right)_{,i}\,, \\
g_{ij}&=&a^2\left[\left(1-2\psi^{(1)}-\psi^{(2)}\right)\delta_{ij}
+2E_{,ij}^{(1)}+E_{,ij}^{(2)}\right]\,.
\end{eqnarray}

The energy-momentum tensor for multiple scalar fields is given by:
\begin{eqnarray}
T_{\mu\nu} &=& \sum_k \left[
\varphi_{k,\mu}\varphi_{k,\nu}
-\frac{1}{2}g_{\mu\nu}g^{\alpha\beta}\varphi_{k,\alpha}
\varphi_{k,\beta}\right]-g_{\mu\nu} V(\{\varphi_i\})\,, \label{enmomtensor} 
\end{eqnarray}
All the variables are split into background plus perturbations:
\begin{equation}
\label{tensor_split}
T= T^{(0)}+T^{(1)}+\frac{1}{2}T^{(2)}+\ldots\,,
\end{equation}
The evolution for background and perturbations is given by the
Einstein equations, $G_{\mu\nu}=8\pi G T_{\mu\nu}$, and the
Klein-Gordon equations for the scalar fields, see \ref{BEP}.

Let us specialize to a case when there is a single scalar field
determining the background evolution, $\varphi$, and the rest have
constant background values $\sigma_i^{(0)\,'}=0$ (usually
$\sigma_i^{(0)}=0$, but this is not necessary). This requires that
$V_{,\sigma_i}=0$ through the Klein-Gordon equation. The scalar fields
are divided as usual into background and perturbations:
\begin{eqnarray}
\varphi = \varphi^{(0)} + \varphi^{(1)} + \frac{1}{2} \varphi^{(2)}\,, \\
\sigma_k = \sigma_k^{(0)} + \sigma_i^{(1)} + \frac{1}{2} \sigma_i^{(2)}\,,
\end{eqnarray}
where $\sigma_i^{(0)}$ is constant and the field fluctuations in the
flat gauge, $\psi^{(1)}=\psi^{(2)}=0$, are Sasaki-Mukhanov variables,
$\varphi^{(1,2)}=Q_{\varphi}^{(1,2)}$ and
$\sigma_i^{(1,2)}=Q_i^{(1,2)}$.

\subsection{Curvature perturbation}
\subsubsection{First order}

The first order curvature perturbation is defined as
\begin{equation}
\zeta^{(1)} = -\psi^{(1)} - {\cal H} \, \frac{\rho^{(1)}}{\rho^{(0)}}\,,
\end{equation}
where $\psi$ is the perturbation of the trace of the spatial metric
and $\rho$ is the total energy density. This combination is gauge
invariant. The evolution of the first order curvature perturbation on
large scales is given by, see for example~\cite{Brandenberger}:
\begin{equation}
\zeta^{(1)\,'} = -\frac{\cal H}{\rho^{(0)} + p^{(0)}} \, \Gamma^{(1)}\,,
\end{equation}
where $\Gamma^{(1)}$ is the iso-curvature perturbation or
non-adiabatic pressure,
\begin{equation}
\Gamma^{(1)} = p^{(1)} - \frac{p^{(0)\,'}}{\rho^{(0)\,'}} \, \rho^{(1)}\,.
\end{equation}
For adiabatic perturbations, $\Gamma^{(1)}=0$, and the curvature
perturbation is conserved on large scales. This happens when there is
a single fluid with a definite equation of state or for one scalar
field on super-Hubble scales. However, in the general case where there
are multiple fluids or scalar fields, the iso-curvature perturbation
does not necessarily vanish.


\subsubsection{Second order}

In the second order a similar equation is obeyed on large
scales,\footnote{This has been generalized to all orders and scales by
Langlois and Vernizzi \cite{Langlois:2005ii}.} see
also~\cite{Malik:2003mv}
\begin{equation}
\zeta^{(2)\,'} = - \frac{\cal H}{\rho^{(0)}+p^{(0)}} \, \Gamma^{(2)}
\end{equation}
where
\begin{eqnarray}
\zeta^{(2)} &=& -\psi^{(2)} - {\cal H} \frac{\rho^{(2)}}{\rho^{(0)\,'}} 
+ 2{\cal H} \frac{\rho^{(1)\,'} \rho^{(1)}}{\left.\rho^{(0)\,'}\right.^2} 
+ 2 \frac{\rho^{(1)}}{\rho^{(0)\,'}} \left( \psi^{(1)\,'} + 2{\cal H} 
\psi^{(1)} \right) \nonumber \\
& & + \left(\frac{\rho^{(1)}}{\rho^{(0)\,'}} \right)^2 \left( {\cal H}' 
+ 2{\cal H}^2 - {\cal H} \frac{\rho^{(0)\,''}}{\rho^{(0)\,'}} \right)
\end{eqnarray}
and
\begin{eqnarray}
\Gamma^{(2)} &=& p^{(2)} - \frac{p^{(0)\,'}}{\rho^{(0)\,'}} \, 
\rho^{(2)} + p^{(0)\,'} \left[ 2 \frac{\rho^{(1)}}{\rho^{(0)\,'}} 
\left( \frac{\rho^{(1)\,'}}{\rho^{(0)\,'}} - \frac{p^{(1)\,'}}{p^{(0)\,'}} 
\right) \right. \nonumber \\
& & \left. + \left(\frac{\rho^{(1)}}{\rho^{(0)\,'}}\right)^2 
\left( \frac{p^{(0)\,''}}{p^{(0)\,'}} - \frac{\rho^{(0)\,''}}{\rho^{(0)\,'}} 
\right) \right] - \frac{2}{\rho^{(0)} + p^{(0)}} \left.\Gamma^{(1)}\right.^2 
+ 4\zeta^{(1)}\Gamma^{(1)}\,.
\end{eqnarray}
Again if isocurvature contribution vanishes then $\zeta^{(2)}$ is
conserved on super-Hubble scales. This has already been shown in
\cite{Malik:2003mv,Vernizzi:2004nc} and it has also been generalized
to all orders \cite{Langlois:2005ii,Lyth:2004gb}. Going to higher
orders gives new possibilities, namely $\zeta^{(1)}$ could be
conserved but $\zeta^{(2)}$ might not, since it may happen that
$\Gamma^{(1)}=0$ but $\Gamma^{(2)}\neq 0$. This is especially
interesting situation for the cosmological non-Gaussianity which is
defined as
\begin{equation}
\label{fnl-def}
\zeta^{(2)} = -\frac{3}{5} \, f_{NL} \left.\zeta^{(1)}\right.^2
\end{equation}
If $\Gamma^{(1)}=0$ then $\zeta^{(1)}$ is conserved but
$\Gamma^{(2)}\neq 0$ then $\zeta^{(2)}$ changes and therefore also the
non-Gaussianity parameter, $f_{NL}$, changes.


\subsection{Evolution of the Perturbations}

From now on we adopt the flat gauge, $\psi_1=\psi_2=0$, and neglect
terms that contain two or more spatial derivatives. We also take only
one of the fields to be evolving at the background and call this
$\varphi$ and the rest of the fields do not evolve at the background
level, $\sigma_i'=0$. For details and general expressions see
\cite{Malik:2005cy}. The components of the Einstein tensor and energy
momentum tensor have been given in \ref{BEP}.

The evolution equations for the background are given by the Friedmann
equation,
\begin{equation}
3M_P^2 {\cal H}^2 = \frac{1}{2} \left.\varphi^{(0)\,'}\right.^2 + 
a^2 V_0\,, \label{friedmann}
\end{equation}
and the Klein-Gordon equation,
\begin{equation}
\varphi^{(0)\,''} + 2{\cal H} \varphi^{(0)\,'} + a^2 V_{,\varphi} 
= 0 \,. \label{kl0}
\end{equation}
In addition one can write the equation for the evolution of the Hubble
parameter, which is not independent of the other two equations:
\begin{equation}
{\cal H}' = {\cal H}^2 - \frac{1}{2M_P^2} \left.\varphi^{(0)\,'}\right.^2\,. 
\label{hdot}
\end{equation}
The Klein-Gordon equations at 1st order read
\begin{eqnarray}
Q^{(1)\,''}_{\varphi} + 2{\cal H} Q^{(1)\,'}_{\varphi} - \varphi^{(0)\,'} 
\phi^{(1)\,'} + a^2 V_{,\varphi\varphi} Q^{(1)}_{\varphi} - 
\triangle Q^{(1)}_{\varphi} = 0 \label{kl1phi}\,, \\
Q^{(1)\,''}_k + 2{\cal H} Q^{(1)\,'}_k + a^2\sum_n V_{,\sigma_k\sigma_n} 
Q^{(1)}_n - \triangle Q^{(1)}_k = 0 \,. \label{kl1sigma}
\end{eqnarray}
We will not need the 2nd order Klein-Gordon equations at all.  This is
due to the fact that on the background level only $\varphi^{(0)}$ has
a non-trivial evolution.

The gauge-invariant curvature perturbation at 1st order in terms of
the Mukhanov variables is given by:
\begin{equation}
\zeta^{(1)} = - \frac{\cal H}{\varphi^{(0)\,'}}\, Q^{(1)}_{\varphi}\,. 
\label{zeta1}
\end{equation}
This is a conserved quantity on large scales (with the approximations
made $\zeta^{(1)\,'}=0$). Using $0-i$ Einstein equation one obtains:
\begin{equation}
Q^{(1)\,'}_{\varphi} = -\left( \frac{3{\cal H}V_0}{\rho^{(0)}} + 
\frac{a^2 V_{,\varphi}}{\varphi^{(0)\,'}} \right) \, 
Q^{(1)}_{\varphi}\,. \label{q1dot}
\end{equation}
The corresponding conserved variable at 2nd order is given
by~\cite{Malik:2005cy}:
\begin{eqnarray}
\zeta^{(2)} &=& \frac{\rho^{(0)}}{3 V_0 \varphi^{(0)\,'}} \, 
Q_{\varphi}^{(2)\,'} + \frac{a^2 \rho^{(0)}}{3V_0 
\left.\varphi^{(0)\,'}\right.^2} \, V_{,\varphi} Q_{\varphi}^{(2)} 
+ \left( \frac{\cal H}{\varphi^{(0)\,'}} Q^{(1)}_{\varphi} \right)^2 
\left[ \frac{3V_0}{\rho^{(0)}} \right. \nonumber \\
& & \left. + \frac{a^4 \rho^{(0)} V_{,\varphi}^2}{3V_0 {\cal H}^2 
\left.\varphi^{(0)\,'}\right.^2} + \frac{a^2 \rho^{(0)} 
V_{,\varphi\varphi}}{3V_0 {\cal H}^2} + \frac{2 V_{,\varphi} 
\varphi^{(0)\,'}}{{\cal H} V_0} + \frac{6 \left.\varphi^{(0)\,'}
\right.^2}{a^2 \rho^{(0)}} - 4 \right] \nonumber \\
& & + \frac{\rho^{(0)}}{3V_0 \left.\varphi^{(0)\,'}\right.^2} 
\sum_k \left.Q_k^{(1)\,'}\right.^2 + \frac{a^2 \rho^{(0)}}{3 V_0 
\left.\varphi^{(0)\,'}\right.^2} \sum_{k,n} V_{,\sigma_k\sigma_n} 
Q_k^{(1)} Q_n^{(1)}\,. \label{zeta2}
\end{eqnarray}
Solving $\phi^{(1)}$ and $\phi^{(2)}$ from $0-0$ Einstein equations
and then taking the divergence of the 2nd order $0-i$ part of the
Einstein equation and then taking the inverse Laplacian, we obtain
\begin{equation}
Q^{(2)\,'}_{\varphi} + \left( \frac{3{\cal H}V_0}{\rho^{(0)}} + 
\frac{a^2 V_{,\varphi}}{\varphi^{(0)\,'}} \right) Q^{(2)}_{\varphi} 
= J_{\varphi}\,, \label{q2}
\end{equation}
where
\begin{eqnarray}
J_{\varphi} &=& -\frac{1}{\varphi^{(0)\,'}} \left\{ 
\left.Q^{(1)\,'}_{\varphi}\right.^2 + \frac{3{\cal H}V_0}{\rho^{(0)}}\, 
Q^{(1)\,'}_{\varphi} Q^{(1)}_{\varphi} + \left.Q^{(1)}_{\varphi}\right.^2 
\left[ a^2 V_{,\varphi\varphi} \right. \right. \nonumber \\
& & \left. + \frac{6a^2{\cal H}V_{,\varphi} \varphi^{(0)\,'}}{\rho^{(0)}} 
+ \frac{9 {\cal H}^2 V_0 \left.\varphi^{(0)\,'}\right.^2}{a^2 
\left.\rho^{(0)}\right.^2} \right] + \sum_k \left.Q^{(1)\,'}_k\right.^2 
\nonumber \\
& & \left. + a^2 \sum_k V_{,kl} Q^{(1)}_k Q^{(1)}_l + 
\frac{6{\cal H} V_0}{\rho^{(0)}} \triangle^{-1} \sum_k 
\left( Q^{(1)\,'}_k Q^{(1)}_{k,j} \right)^{,j} \right\}\,. \label{j2}
\end{eqnarray}
Then the solution to Eq. (\ref{q2}) is given by:
\begin{equation}
Q^{(2)}_{\varphi} = C\, \frac{\varphi^{(0)\,'}}{\cal H} 
+ \frac{\varphi^{(0)\,'}}{\cal H} \int^{\eta} 
d\eta \frac{\cal H}{\varphi^{(0)\,'}} \, J_{\varphi}\,. \label{solq2}
\end{equation}
It would seem that the integral over the source becomes ill-defined at
times when $\varphi^{(0)\,'}=0$, which happens during
oscillations. During inflation the background field is slow-rolling
and this problem is circumvented but after inflation when the inflaton
oscillates before decaying it is a problem. Since $Q^{(1)}_i$ and
their derivatives are well defined variables with no singularities,
the Klein-Gordon equations guarantee this, the only singularity in the
source integral comes from $\left.1/\varphi^{(0)\,'}\right.^{2}$ which
gets multiplied by $\varphi^{(0)\,'}$ after integration. We can
extract that singular behavior by expanding all the quantities around
the time $\eta_i$, where $\varphi^{(0)\,'}(\eta_i)=0$. Then
\begin{equation}
\varphi^{(0)\,'} = \varphi^{(0)\,''} (\eta_i) \, (\eta-\eta_i) + 
{\cal O}(\eta-\eta_i)^2 +...\,, \label{vel1}
\end{equation}
where $\varphi^{(0)\,''}(\eta_i)\neq 0$ during oscillations and the
rest of the terms inside the integral are also non-vanishing, but
regular at $\eta=\eta_i$, so we can replace them in the most singular
term by constants. Then
\begin{equation}
\frac{\varphi^{(0)\,'}}{\cal H} \int^{\eta} d\eta \frac{1}
{\varphi^{(0)\,'}} \, J_{\varphi} \sim (\eta-\eta_i) \int^{\eta} 
d\eta \frac{1}{(\eta-\eta_i)^2} \sim const. \label{source}
\end{equation}
when $\eta\to\eta_i$ so that $Q^{(2)}_{\varphi}$ is completely regular
even during oscillations. Next one has to define the constant $C$ in
Eq.~(\ref{solq2}) in terms of some initial condition at
$\eta=\eta_0$. Here we have to choose a time such that
$\varphi^{(0)\,'}(\eta_0)\neq 0$, since otherwise the homogeneous term
vanishes and we are not able to define the constant. Then
\begin{equation}
Q^{(2)}_{\varphi} = \frac{\varphi^{(0)\,'}}{\varphi^{(0)\,'}(\eta_0)} \, 
\frac{{\cal H}_0}{\cal H} \, Q^{(2)}_{\varphi}(\eta_0) + 
\frac{\varphi^{(0)\,'}}{\cal H} \int_{\eta_0}^{\eta} d\eta 
\frac{\cal H}{\varphi^{(0)\,'}} \, J_{\varphi}\,. \label{sol2q2}
\end{equation}
Now we are solving a definite integral over the source term so that
there can be several points along the integration where
$\varphi^{(0)\,'}=0$, which upon multiplication will not be removed,
since it is done only at the final time. This is not really dangerous
if we are able to do the integral in closed form but if one needs to
evaluate it numerically it is a real complication, see \ref{TAPTZ} for
a toy example that clarifies the complications related to the
oscillations.

Finally, what we are really interested in is the curvature
perturbation, $\zeta^{(2)}$, and not the Sasaki-Mukhanov variable
$Q^{(2)}_{\varphi}$. However, the curvature perturbation is not well
defined during oscillations since it is obtained from the
Sasaki-Mukhanov variable by dividing with $\varphi^{(0)\,'}$, see
Eq. (\ref{zeta2}). This problem occurs even in first order, see
Eq. (\ref{zeta1}). The solution is to consider instead
$\left.\varphi^{(0)\,'}\right.^2 \zeta$ or $(1+w)\zeta$, where
$1+w=\left.\varphi^{(0)\,'}\right.^2/(a^2\rho^{(0)})$, which has been
used in first order calculations successfully and it also works in
second order. Using Eqs.~(\ref{zeta2},\ref{q2},\ref{solq2})
$\zeta^{(2)}$ can be rewritten as
\begin{eqnarray}
\zeta^{(2)} &=& C - \frac{2{\cal H}}{\left.\varphi^{(0)\,'}\right.^2} 
\triangle^{-1} \left( \sum_k Q^{(1)\,'}_k Q^{(1)}_{k,j} \right)^{,j} 
+ \int^{\eta} d\eta \frac{\cal H}{\left.\varphi^{(0)\,'}\right.^2}
\left[ \sum_k \left. Q^{(1)\,'}_k \right.^2 \right. \nonumber \\
& & \left. + a^2 \sum_{k,n} V_{,\sigma_k\sigma_n} Q^{(1)}_k Q^{(1)}_n 
+ \frac{6{\cal H} V_0}{\rho^{(0)}} \, \triangle^{-1} 
\left( \sum_k Q^{(1)\,'}_k Q^{(1)}_{k,j} \right)^{,j} \right] \,.
\label{solzeta2}
\end{eqnarray}
Similar reasoning as we have given for $Q^{(2)}_{\varphi}$ in
Eqs.~(\ref{vel1},\ref{source})) onwards shows that
$\left.\varphi^{(0)\,'}\right.^2 \zeta^{(2)}$ is well defined. And
actually for points where $\varphi^{(0)\,'}=0$ it is given by the
second term in Eq. (\ref{solzeta2}) since the integral term
vanishes. Another way to derive the result for the case of
$\varphi^{(0)\,'}=0$ is to use it in the original Einstein equations,
in which case one can solve the Mukhanov variable directly at that
particular time and obtain the result given by the second term in
Eq. (\ref{solzeta2}). The non-local term can be rewritten as (see
\ref{NLT})
\begin{equation}
\triangle^{-1} \left( \sum_k Q^{(1)\,'}_k Q^{(1)}_{k,j} \right)^{,j} 
= \frac{1}{2} \sum_k Q^{(1)\,'}_k Q^{(1)}_k
\end{equation}
The final formula for second order curvature perturbation is
\begin{eqnarray}
\zeta^{(2)} &=& C - \frac{{\cal H}}{\left.\varphi^{(0)\,'}\right.^2} \sum_k
Q^{(1)\,'}_k Q^{(1)}_k + \int^{\eta} d\eta \frac{\cal
  H}{\left.\varphi^{(0)\,'}\right.^2} \left[ \sum_k \left. Q^{(1)\,'}_k
  \right.^2 \right. \nonumber \\
& & \left. + a^2 \sum_{k,n} V_{,\sigma_k\sigma_n} Q^{(1)}_k Q^{(1)}_n 
+ \frac{3{\cal H} V_0}{\rho^{(0)}} \, \sum_k Q^{(1)\,'}_k Q^{(1)}_k \right] \,.
\label{solzeta2fin}
\end{eqnarray}
Note that the above expression does not contain any non-local
term~\footnote{Our expression improves the one obtained in
Ref.~\cite{Malik:2005cy}, since we got rid off the non-local
contributions by solving them in terms of local contributions, see
\ref{NLT}.}


\section{Massless preheating}

As an application to our formalism, let us consider a simple
potential,
\begin{equation}
V(\varphi,\sigma) = \frac{1}{4} \lambda \varphi^4 + \frac{1}{2} g^2 \varphi^2
\sigma^2\,, \label{potential}
\end{equation}
where $\varphi$ is the inflaton and $\sigma$ field is coupled to the
inflaton with a coupling strength $g$. After inflation the inflaton,
$\varphi$, begins coherent oscillations with an initial amplitude
$\sim {\cal O}(1)M_{P}$. During this process it amplifies $\sigma_k$
non-thermally. This happens because $\sigma$ field sees the time
varying mass due to its coupling to the inflaton. In this paper we are
interested in studying the fluctuations of the low momentum modes of
the $\sigma$ field. This has been studied in first order perturbation theory in several papers \cite{Greene:1997fu,Kaiser:1997hg,Bassett:1999cg}.

For the above setup the Friedmann equation is given by:
\begin{equation}
3 M_P^2 {\cal H}^2 = \frac{1}{2} \varphi'^2 + 
\frac{1}{4} a^2 \lambda \varphi^4\,,
\label{friedmann2}
\end{equation}
here we consider the inflaton oscillations is the main source for
expanding the Universe, and the Klein-Gordon equation:
\begin{equation}
\varphi'' + 2{\cal H} \varphi' + a^2 \lambda \varphi^3 = 0\,. 
\label{kleingordon}
\end{equation}
Note that for the purpose of illustration we assume the VEV of
$\sigma=0$~\footnote{At initial stages $\sigma=0$ is a very good
approximation, as the VEV of $\varphi$ decreases it eventually sets a
motion for $\sigma$ field due to the backreaction of produced $\sigma$
quanta. In Ref.~\cite{Nambu} the authors analyze non-linear
perturbations when the backreaction is important, however they do not
estimate the non-Gaussianity parameter in their case. In our case we
estimate non-Gaussianity from the exponential growth of the linear
perturbations of $\sigma$ field.}.  In the calculation it is useful to
scale out the expansion by $\varphi=\chi/a$ leading to
\begin{eqnarray}
& 3 M_P^2 {\cal H}^2 = \frac{1}{2a^2} \left(\chi' - {\cal H} \chi \right)^2 +
\frac{1}{4a^2} \lambda \chi^4\,, \label{refriedmann} \\
& \chi'' + \lambda \chi^3 + \frac{a''}{a} \, \chi = 0\,. \label{reklein}
\end{eqnarray}
A simple solution can be obtained for $\chi$ during oscillations while
neglecting the expansion rate in Eq. (\ref{refriedmann}), for details
see \cite{Greene:1997fu,Kaiser:1997hg}:
\begin{equation}
\chi(\eta) = \bar{\chi}\, cn\left(x-x_0,\frac{1}{\sqrt{2}}\right), \qquad
x=\sqrt{\lambda} \bar{\chi} \,\eta \,,
\label{chisol}
\end{equation}
where $\bar{\chi}$ is the amplitude of the oscillations, $cn$ is a
Jacobian elliptic function and $x_0=2.44$ matches the solution to the
slow-roll solution at the end of inflation. The period of oscillations
is $T=\Gamma(1/4)^2/\sqrt{\pi}\approx 7.4$. Then
Eq. (\ref{refriedmann}) can be rewritten as:
\begin{equation}
3 M_P^2 a'^2 = \frac{1}{2} \chi'^2 + \frac{1}{4} \lambda \chi^4 = \frac{1}{4}
\lambda \bar{\chi}^4\,, \label{refried}
\end{equation}
where in the last equality the fact that energy of $\chi$ is conserved
when expansion is neglected as guaranteed by Eq. (\ref{reklein}). The
solution of Eq. (\ref{refried}) gives $a\sim \eta$ from which follows
that $a''=0$, so that the approximation is consistent. This also
results into ${\cal H}=1/\eta$.

Now we study the gauge invariant fluctuations of $\sigma$, which is
described by the Mukhnaov variable $Q^{(1)}_{\sigma}$. However in this
particular case the Mukhanov variable is equivalent to the field
fluctuations, $\delta\sigma_{k}$, because the background value of
$\sigma$ is vanishing. Note that although $\sigma$ field carries
isocurvature fluctuations, it does not source the first order
curvature perturbations due to its vanishing VEV. Therefore in our
case the fluctuations in $\sigma$ cannot be constrained by the current
constraints from the isocurvature perturbation. However these
fluctuations do source the second order curvature perturbation and
therefore are subject to the constraints from non-Gaussianity.

The fluctuations of $\sigma$ can be solved from the perturbed
Klein-Gordon equation:
\begin{equation}
Q^{(1)\,''}_{\sigma} + 2{\cal H} Q^{(1)\,'}_{\sigma} 
+ a^2 g^2 \left.\varphi^{(0)}\right.^2 Q^{(1)}_{\sigma} 
- \triangle Q^{(1)}_{\sigma} = 0 \label{q1sigma}
\end{equation}
It is again useful to make the re-scaling $Q^{(1)}_{\sigma}=
\chi^{(1)}_{\sigma}/a$ and change of variable
$x=\sqrt{\lambda}\bar{\chi} \eta$. Fourier transforming
Eq. (\ref{q1sigma}), we obtain:
\begin{equation}
\ddot{\chi}^{(1)}_{\sigma {\bf k}} + \left[\kappa^2 
+ \frac{g^2}{\lambda} cn^2(x-x_0,1/\sqrt{2}) \right] 
\chi^{(1)}_{\sigma {\bf k}} = 0\,, \label{chi1sigma}
\end{equation}
where $\dot{}= d/dx$ and $\kappa^2 = k^2/\lambda \bar{\chi}^2$. Since
for observable scales the dimensionless momentum is small, for example
for length scale of $\sim$~1Mpc it is $\kappa\sim 10^{-50}$, it is
important that the instability band contains also the $k=0$ mode,
which is actually true for Eq. (\ref{chi1sigma}),
see~\cite{Greene:1997fu}. This also justifies taking $\kappa=0$ while
estimating analytically the non-Gaussianity parameter, for a detailed
discussion also see \ref{SHM}~\footnote{The Floquet index defined in
Eq.~(\ref{result1}) has a mild scale dependence for cosmologically
interesting scales. }.

Cosmologically the most interesting solutions are the ones with $1\leq
g^2/\lambda \leq 3$, since the condition for $\sigma$ to be a light
field during inflation is that is mass is less than the Hubble
parameter, more specifically $g\varphi^{(0)} < (3/2) H$. Using the
Friedmann equation this can be written as $g^2/\lambda < 3/16
(\varphi^{(0)}/M_P)^2$. During the slow-roll phase $\epsilon = M_P^2/2
\, (V_{,\varphi}/V_0)^2 = (M_P/(2\sqrt{2}\varphi^{(0)}))^2 < 1$, so
that during slow-roll the condition is obeyed for $1<g^2/\lambda<3$.
This also guarantees that the primordial fluctuations (generated
during inflation) of $\sigma$ field are scale invariant.

The second order curvature perturbation can be calculated if in
Eq. (\ref{solzeta2}) we approximate the
$\left.\varphi^{(0)\,'}\right.^2$ with its time average
$(1+w)a^2\rho^{(0)}$, where $w=1/3$ for radiation dominated
universe. Then
\begin{eqnarray}
\zeta^{(2)} &=& \zeta^{(2)}(\eta_0) + \frac{{\cal H}_0}
{(1+w)a_0^2 \rho^{(0)}(\eta_0)} \, Q^{(1)\,'}_{\sigma}
(\eta_0) Q^{(1)}_{\sigma}(\eta_0) - \frac{\cal H}
{(1+w)a^2 \rho^{(0)}} Q^{(1)\,'}_{\sigma} Q^{(1)}_{\sigma} \nonumber \\
&+& \int_{\eta_0}^{\eta} d\eta \frac{\cal H}{(1+w)a^2\rho^{(0)}} 
\left\{ \left.Q^{(1)\,'}_{\sigma}\right.^2 + a^2g^2 
\left.\varphi^{(0)}\right.^2 \left.Q^{(1)}_{\sigma}\right.^2 
+ \frac{3{\cal H} U_0}{\rho^{(0)}} Q^{(1)\,'}_{\sigma} 
Q^{(1)}_{\sigma} \right\}\,. \label{zeta2chi}
\end{eqnarray}
We simplify the analysis further by disregarding the oscillatory
behavior inside the integral, since the first two terms are positive
definite quantities there is no problem with this and the last term in
the integral has an extra Hubble parameter in it so that it decays
away faster than the first two ones and is therefore less
significant. Once the amplitude of oscillation is known (although now
the amplitude grows exponentially) the oscillations can be averaged by
adding a factor of $\sim 1/2$ which corresponds to a time average of a
square of an oscillating function. We also estimate the real solution
Eqs.~(\ref{chi1sol}-\ref{tildet}), see \ref{SHM}, by an exponentially
growing function
\begin{equation}
\label{result1}
\chi^{(1)}_{\sigma} \approx \frac{1}{2} \chi^{(1)}_{\sigma}(\eta_0) 
e^{\mu(s)\,(x-x_0)}\,,
\end{equation}
where the $1/2$ is due to averaging over the oscillations and $\mu(s)$
is given by Eq. (\ref{mus}), where $s=g^2/\lambda$. The growth index
$\mu(s)$ gets values between $0$ and $\mu_{\max}\approx 0.2378$. Since
the expansion has been neglected after re-scaling of the field, we
take only the leading order contributions in terms of the expansion in
Eq. (\ref{zeta2chi}). After simplifying the resulting expression we
get
\begin{eqnarray}\label{fin1}
\zeta^{(2)} = \zeta^{(2)}(\eta_0) + \frac{1}{4} r^2 \left.
\zeta^{(1)}\right.^2 \left[ \frac{1}{x^2} \, e^{2\mu(s) (x-x_0)} - 
\frac{1}{x_0^2} + \left( \frac{s}{2} - \mu(s)^2 \right) 
\int_{x_0}^x dx \frac{1}{x} \, e^{2\mu(s) (x-x_0)} \right] \nonumber \\
\end{eqnarray}
where $r=Q^{(1)}_{\sigma}(\eta_0)/Q^{(1)}_{\varphi}(\eta_0)$ is the
relative magnitude of the $\sigma$ and $\varphi$ perturbations at the
end of inflation. If both fields are light one expects $r\approx
1$. In reality $\sigma$ becomes massive during or after the end of
slow-roll period, so that there is a slight damping. Second the
perturbations can have different slow-roll parameters so that $r$ can
be mildly scale dependent.

\begin{figure}
\begin{center}
\includegraphics[width=0.75\textwidth]{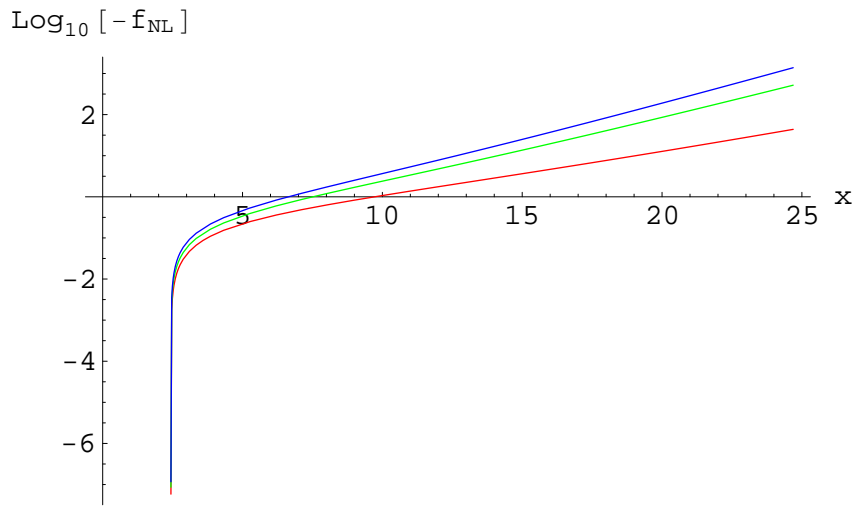}
\caption{\small The evolution of the non-Gaussianity parameter,
$f_{NL}$, with different values of $g^2/\lambda=1.2,\,1.5,\,1.875$
with red, green and blue lines. $g^2/\lambda=1.875$ is the
case with largest characteristic exponent. The end point corresponds
to 3 oscillations of the inflaton and at that point the largest
$-f_{NL}=1382$.}\label{lphi4f1}
\end{center}
\end{figure}

\begin{figure}
\begin{center}
\includegraphics[width=0.75\textwidth]{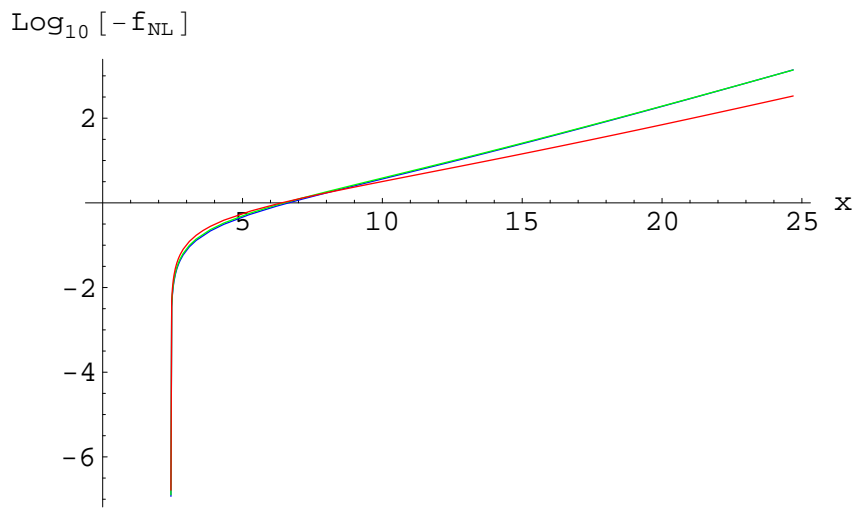}
\caption{\small The same as previous figure with
$g^2/\lambda=1.875,\,2,\,2.5$ with blue, green and red lines. The
cases with $g^2/\lambda=1.875$ (blue) and $2$ (green) produce almost
identical results, which is not surprising since $\mu_{1.875}\approx
0.238$ and $\mu_2\approx 0.236$.}\label{lphi4f2}
\end{center}
\end{figure}

We illustrate the $f_{NL}$ parameter defined in Eq.~(\ref{fnl-def})
numerically for $x=25$ which corresponds to roughly $3$ inflaton
oscillations. In Plots~[\ref{lphi4f1},\ref{lphi4f2}], we have drawn
$\log[f_{NL}]$ with respect to $x$. Note that the $f_{NL}$ rises
exponentially owing to Eq.~(\ref{fin1}). This is because the mode,
$\chi_{\sigma}^{(1)}$, grows exponentially, i.e., see the discussion
above Eq.~(\ref{growth}).  The same mode sources the second order
curvature perturbation, Eq.~(\ref{fin1}), but not the first order, see
Eq.~(\ref{zeta1}), therefore the $f_{NL}$ parameter is alarmingly
large.

This result exceeds the current bound on non-Gaussianity, $-58\leq
f_{NL}\leq 134$ set by WMAP~\cite{WMAP}. Our results show that
massless preheating on the parameter range, $1<g^2/\lambda < 3$, is
already ruled out. We do not expect that our result will be modified
dramatically even if the backreaction due to produced $\sigma$ quanta
is taken into account.

Note that the maximal growth index is obtained at
$g^2/\lambda=15/8=1.875$ and not $g^2/\lambda=2$ as has usually been
assumed in the literature, see for instance the recent
paper~\cite{Nambu}. However the difference between the two cases is
negligible as is shown is Plot~\ref{lphi4f2}.

%
%
%
\section{Conclusion}

Following our previous claim~\cite{asko1}, we pointed out in this
paper that preheating in general can produce very large primordial
non-Gaussianity.  We substantiate our point by providing a formalism
which can take into account of the fluctuations for more than one
field, therefore accounting for the isocurvature fluctuations which can
source the first and second order curvature perturbations. The final
expression for the second order curvature perturbation is given by
Eq.~(\ref{solzeta2fin}).

We studied a particular case of massless preheating as an example,
where the second order curvature perturbations are fed by the the
isocurvature fluctuations but not the first order.  This gives rise to
a large non-Gaussianity which already exceeds the currently observed
limit. The non-Gaussianity parameter grows exponentially and the final
number depends on the number of oscillations before the backreaction
kicks in to halt preheating completely, see the
Plots~[\ref{lphi4f1},\ref{lphi4f2}].  Based on our result we rule out
completely massless preheating on the parameter range, $1<g^2/\lambda
< 3$.

Our results are encouraging and obviously demands thorough study of
non-Gaussianity generated during preheating. One of the goal will be
to setup a numerical code where more than one field can participate in
the expansion of the Universe. This would definitely improve the
non-Gaussianity bound set in the current paper.

\section{Acknowledgments}

The authors are thankful to Robert Brandenberger, Kari Enqvist, Lev
Kofman, Andrew Liddle, David Lyth, and Antti V\"aihk\"onen. The authors
also acknowledge the hospitality of the ASPEN physics center where
part of the work was carried out during SUPERCOSMOLOGY workshop.

\appendix \setcounter{section}{0} \setcounter{equation}{0}

\section{Appendix}
\subsection{Setting the background equations and perturbations}\label{BEP}

The background Einstein tensor is
\begin{eqnarray}
G^{(0)0}_{\phantom{(0)0}0} &=& -\frac{3}{a^2}\frac{a'^2}{a^2}\,, \\
G^{(0)i}_{\phantom{(0)i}j} &=& \frac{1}{a^2}\left[\frac{a'^2}{a^2}-
2\frac{a''}{a}\right]\,,
\end{eqnarray}
The Einstein tensor at first order is given by:
\begin{eqnarray}
\label{G00_1}
G^{(1)0}_{\phantom{(1)0}0} &=& 
\frac{6}{a^2}\,\frac{a'^2}{a^2}\phi^{(1)} \,, \\
\label{G0i_1}
G^{(1)0}_{\phantom{(1)0}i} &=& 
-\frac{2}{a^2}\,\frac{a'}{a}\phi_{,i}^{(1)}
\\
G^{(1)i}_{\phantom{(1)i}j} &=& 
\frac{1}{a^2}\left[
4\frac{a''}{a}\phi^{(1)} + 2\left(\frac{a'}{a}
\phi^{(1)\,'}-\frac{a'^2}{a^2}\phi^{(1)}\right)
\right] \delta^i_{~j} \,,
\end{eqnarray}
and at second order
\begin{eqnarray}
\label{G0i_2}
G^{(2)0}_{\phantom{(2)0}0} &=& 
-\frac{6}{a^2}\left[-\frac{a'^2}{a^2}\phi^{(2)} + 4\frac{a'^2}{a^2} 
\left.\phi^{(1)}\right.^2 \right] \,, \\
\label{G00_2}
G^{(2)0}_{\phantom{(2)0}i} &=& -\frac{2}{a^2}\left[
\frac{a'}{a}\phi^{(2)} - 4\frac{a'}{a} \left.\phi^{(1)}\right.^2
\right]_{,i} \,,
\\
\label{Gij_2}
G^{(2)i}_{\phantom{(2)i}j} &=& \frac{2}{a^2}\left[
-8\left(\frac{a'}{a}\phi^{(1)}\phi^{(1)\,'}+\frac{a''}{a} 
\left.\phi^{(1)}\right.^2\right) + \frac{a'}{a} \phi^{(2)\,'} \right. 
\nonumber \\
& & \left. +4\frac{a'^2}{a^2} \left.\phi^{(1)}\right.^2 
+ 2\frac{a''}{a}\phi^{(2)} - \frac{a'^2}{a^2}\phi^{(2)} \right] \delta^i_{~j}
\,,
\end{eqnarray}
Then the energy-momentum tensor of the background is given by:
\begin{eqnarray}
T_{\phantom{(0)0}0}^{(0)0} = -\frac{1}{2a^2} \left.
\varphi^{(0)\,'}\right.^2 - V_0 \\
T_{\phantom{(0)i}j}^{(0)i} = \left(\frac{1}{2a^2} \left.
\varphi^{(0)\,'}\right.^2 - V_0
\right) \delta^i_{\phantom{i}j}
\end{eqnarray}
At first order in perturbations
\begin{eqnarray}
\label{deltaT100}
T^{(1)0}_{\phantom{(1)0}0} &=& -\frac{1}{a^2}\left(
\varphi^{(0)\,'} Q_{\varphi}^{(1)\,'}-\left.
\varphi^{(0)\,'}\right.^2 \phi^{(1)}
\right)-V_1   \,, \\
\label{deltaT10i}
T^{(1)0}_{\phantom{(1)0}i} 
&=& -\frac{1}{a^2}\varphi^{(0)\,'} Q_{\varphi,i}^{(1)}\,, \\
T^{(1)i}_{\phantom{(1)i}j} &=& \frac{1}{a^2}\left[
\left(\varphi^{(0)\,'} Q_{\varphi}^{(1)\,'} - \left.\varphi^{(0)\,'}
\right.^2\phi^{(1)}
\right) - a^2 V_1 \right]\delta^i_{~j}\,,
\end{eqnarray}
and at second order
\begin{eqnarray}
T^{(2)0}_{\phantom{(2)0}0} &=& -\frac{1}{a^2}\left[
\varphi^{(0)\,'} Q_{\varphi}^{(2)\,'} - 4\varphi^{(0)\,'} \phi^{(1)}
Q_{\varphi}^{(1)\,'} - \left.\varphi^{(0)\,'}\right.^2 \phi^{(2)} 
\right. \nonumber \\
& & \left. + 4 \left.\varphi^{(0)\,'}\right.^2 \left.\phi^{(1)}\right.^2 
+ \left.Q_{\varphi}^{(1)\,'}\right.^2+ \sum_k \left.Q_{k}^{(1)\,'}\right.^2
+a^2 V_2 \right] \,, \\
\label{deltaT20i}
T^{(2)0}_{\phantom{(2)0}i} 
&=& -\frac{1}{a^2}\left(\varphi^{(0)\,'} Q_{\varphi,i}^{(2)}
-4 \phi^{(1)} \varphi^{(0)\,'} Q_{\varphi,i}^{(1)} + 2Q_{\varphi}^{(1)\,'}
Q_{\varphi,i}^{(1)} + 2\sum_k Q_k^{(1)\,'} Q_{k,i}^{(1)} \right)\,, \\
T^{(2)i}_{\phantom{(2)i}j} 
&=& \frac{1}{a^2}\left[ \varphi^{(0)'} Q_{\varphi}^{(2)\,'} -
  4\varphi^{(0)\,'} \phi^{(1)} Q_{\varphi}^{(1)\,'} - 
\left.\varphi^{(0)\,'}\right.^2
  \phi^{(2)} \right. \nonumber \\
& & \left. + 4\left.\varphi^{(0)\,'}\right.^2 
\left.\phi^{(1)}\right.^2 + \left.Q_{\varphi}^{(1)\,'}\right.^2
  + \sum_k \left.Q_k^{(1)\,'}\right.^2 - a^2V_2\right]\delta^i_{~j} \,,
\end{eqnarray}
where
\begin{eqnarray}
V_0 &=& V(\sigma_k=\sigma_k^{(0)}) \\
V_1 &=& V_{,\varphi} Q_{\varphi}^{(1)} \\
V_2 &=& V_{,\varphi} Q_{\varphi}^{(2)} + V_{,\varphi\varphi}
(Q_{\varphi}^{(1)})^2 + \sum_{m,n} V_{,\sigma_m\sigma_n} 
Q_m^{(1)} Q_n^{(1)}\,.
\end{eqnarray}


\subsection{Huh, getting rid of the non-local terms!}\label{NLT}

The non-local term is a complication that every one would love to get
rid off. Here we work more generally and allow all the scalar fields
to have non-vanishing background values. Then the Klein-Gordon equations
on large scales (neglecting the gradient term) are:
\begin{eqnarray}
Q^{(1)\,''}_k + 2{\cal H} Q^{(1)\,'}_k + \sum_l \left[ a^2 V_{,kl} 
- \frac{1}{M_{pl}^2 a^2} \left(\frac{a^2 \varphi^{(0)\,'}_k 
\varphi^{(0)\,'}_l}{\cal H} \right)' \right] Q^{(1)}_k = 0\,.
\end{eqnarray}
Now we multiply the Klein-Gordon equation with $\partial_j Q_k$, take
the divergence of the resulting equation, sum over all fields and
operate on it by $\triangle^{-1}$. The result is
\begin{eqnarray}
\triangle^{-1} \sum_k \left(Q^{(1)\,''}_k Q^{(1)}_{k,j} \right)^{,j} 
+ 2{\cal H} \triangle^{-1} \sum_k \left(Q^{(1)\,'}_k Q^{(1)}_{k,j} 
\right)^{,j} & & \nonumber \\
+ \triangle^{-1} \sum_{k,l} \left[ a^2 V_{,kl} - \frac{1}{M_{pl}^2 a^2} 
\left(\frac{a^2 \varphi^{(0)\,'}_k \varphi^{(0)\,'}_l}{\cal H} \right)' 
\right] \left( Q^{(1)}_{k,j} 
Q^{(1)}_l \right)^{,j} &=& 0\,.
\end{eqnarray}
Next we re-write the first term
\begin{eqnarray}
& \left[\triangle^{-1} \sum_k \left( Q^{(1)\,'}_k Q^{(1)}_{k,j} 
\right)^{,j} \right]' - \triangle^{-1} \sum_k \left( Q^{(1)\,'}_k 
Q^{(1)\,'}_{k,j} \right)^{,j} + 2{\cal H} \triangle^{-1} \sum_k \left(Q^{(1)\,'}_k Q^{(1)}_{k,j} \right)^{,j} \nonumber \\
& + \triangle^{-1} \sum_{k,l} \left[ a^2 V_{,kl} 
- \frac{1}{M_{pl}^2 a^2} \left(\frac{a^2 \varphi^{(0)\,'}_k 
\varphi^{(0)\,'}_l}{\cal H} \right)' \right] 
\left( Q^{(1)}_{k,j} Q^{(1)}_l \right)^{,j} = 0\,.
\end{eqnarray}
In the second term the spatial derivatives and inverse Laplacian can
now be solved (as it can be done also in the last term) since the
potential is symmetric in its derivatives and the same is true of the
extra mass term
\begin{eqnarray}
& \left[\triangle^{-1} \sum_k \left( Q^{(1)\,'}_k Q^{(1)}_{k,j} 
\right)^{,j} \right]' - \frac{1}{2} \sum_k \left(Q^{(1)\,'}_k 
\right)^2 + 2{\cal H} \triangle^{-1} \sum_k 
\left(Q^{(1)\,'}_k Q^{(1)}_{k,j} \right)^{,j} \nonumber \\
& + \frac{1}{2} \sum_{k,l}\left[a^2 V_{,kl}- \frac{1}{M_{pl}^2 a^2} 
\left(\frac{a^2 \varphi^{(0)\,'}_k \varphi^{(0)\,'}_l}{\cal H} 
\right)' \right] Q^{(1)}_k Q^{(1)}_l = 0\,.
\end{eqnarray}
Next rewrite the second term again:
\begin{eqnarray}
\left[\triangle^{-1} \sum_k \left( Q^{(1)\,'}_k Q^{(1)}_{k,j} 
\right)^{,j} \right]' - \frac{1}{2} \left(\sum_k Q^{(1)\,'}_k 
Q^{(1)}_k \right)' + \frac{1}{2} \sum_k Q^{(1)\,''}_k Q^{(1)}_k & & 
\nonumber \\
+ 2{\cal H} \triangle^{-1} \sum_k \left(Q^{(1)\,'}_k Q^{(1)}_{k,j} 
\right)^{,j} + \frac{1}{2} \sum_{k,l}\left[a^2 V_{,kl} 
- \frac{1}{M_{pl}^2 a^2} \left(\frac{a^2 \varphi^{(0)\,'}_k 
\varphi^{(0)\,'}_l}{\cal H} \right)' \right] Q^{(1)}_k Q^{(1)}_l &=& 0\,,
\nonumber \\
\end{eqnarray}
and then use the first order Klein-Gordon equation to rewrite it as
\begin{eqnarray}
\left[\triangle^{-1} \sum_k \left( Q^{(1)\,'}_k Q^{(1)}_{k,j} 
\right)^{,j} \right]' + 2{\cal H} \triangle^{-1} \sum_k 
\left(Q^{(1)\,'}_k Q^{(1)}_{k,j} \right)^{,j} & & \nonumber \\
- \frac{1}{2} \left[\left(\sum_k Q^{(1)\,'}_k Q^{(1)}_k \right)' + 
2{\cal H} \sum_k Q^{(1)\,'}_k Q^{(1)}_k \right] &=& 0\,.
\end{eqnarray}
This can be integrated to yield:
\begin{equation}
\triangle^{-1} \sum_k \left( Q^{(1)\,'}_k Q^{(1)}_{k,j} 
\right)^{,j} = \frac{1}{2} \sum_k Q^{(1)\,'}_k Q^{(1)}_k 
+ \frac{f({\bf x})}{a^2}\,, \label{correspondence}
\end{equation}
where $f({\bf x})$ is a function of space only. For one field
Eq.~(\ref{q1dot}) shows that $f=0$. This is not surprising since for
one dimensional system the first derivative is always proportional to
the function itself where the proportionality depends only on time as
long as we work with the leading order of gradient expansion. For
multiple fields it is enough if the matrix of proportionality is
symmetric and depends only on time. Special case of this is the one
where multiple fields decouple into a system of one field
models. During slow-roll inflation the potential derivatives are of
order the slow-roll parameters~\footnote{Same would be true for the
extra mass term in the general case.}. Then the Klein-Gordon equation
at zeroth order of slow-roll is:
\begin{equation}
Q^{(1)\,''}_{i{\bf k}} + 2{\cal H} Q^{(1)\,'}_{i{\bf k}} 
+ k^2 Q^{(1)}_{i{\bf k}} = 0\,,
\end{equation}
with the well-known solution (during inflation $a=-1/(H\eta)$ with
$H=const.$)
\begin{equation}
Q^{(1)}_{i{\bf k}} = \frac{1}{a\sqrt{2k}} \, 
e^{-ik\eta} \left( 1 + \frac{i}{k\eta} \right)\,.
\end{equation}
The derivative of this solution is proportional to the solution itself
with the proportionality given by a time-dependent function at the
leading order in large scales, $Q^{(1)\,'}_{i{\bf k}}=f_i(\eta)
Q^{(1)}_{i{\bf k}}$ (actually in this case the time derivative
vanishes in leading order but in general it is proportional to
slow-roll parameters, however the main point is that the
proportionality factor in leading order does not depend on ${\bf
k}$). Then the non-local term can be solved using the
Fourier-transform
\begin{eqnarray}
\sum_m \triangle^{-1} \left(Q^{(1)\,'}_m Q^{(1)}_{m,j} 
\right)^{,j} = \sum_m \frac{1}{(2\pi)^3} {\bf k}^{-2} 
\int d^3{\bf k'} Q^{(1)\,'}_{m{\bf k'}} {\bf k}\cdot 
({\bf k}-{\bf k'}) Q^{(1)}_{m({\bf k}-{\bf k'})} = \nonumber \\
\sum_m f_m(\eta) \frac{1}{(2\pi)^3} {\bf k}^{-2} 
\int d^3{\bf k'} Q^{(1)}_{m{\bf k'}} {\bf k}\cdot 
({\bf k}-{\bf k'}) Q^{(1)}_{m({\bf k}-{\bf k'})} = \nonumber \\
\sum_m f_m(\eta) \triangle^{-1} 
\left( Q^{(1)}_m Q^{(1)}_{m,j} \right)^{,j} = 
\frac{1}{2} \sum_m Q^{(1)\,'}_m Q^{(1)}_m\,.
\end{eqnarray}
so that the unknown function in Eq. (\ref{correspondence}) actually
vanishes $f({\bf x})=0$. This result is also valid after inflation
since one only needs to specify correspondence of non-local term to a
local term at one time. In a sense this is just a result of the fact
that after horizon crossing the modes evolve in the same way. Another
way of getting the same result would be to use slow-roll to neglect
the second order time derivatives in the Klein-Gordon equation. Then
the first order time derivative of the Mukhanov variables would be
proportional to a time dependent (but space independent) symmetric
matrix times the Mukhanov variables. Then one can replace the time
derivative in the non-local term and solve it completely to obtain the
same result. One should note that both derivations are only valid in
large scales since the spatial gradient terms were neglected in the
derivations. Therefore for small scales the scale dependence should be
more important so it is not surprising that there it may not be
possible to remove the spatial derivatives.


\subsection{Time averaging and passing through zeros}\label{TAPTZ}

Let us study a similar situation with an example concerning elementary
functions where the integrals can be done explicitly. We take
\begin{equation}
f(x) = C\,\cos(x) + \sin(x) = \cos(x) \left[ C + \int^x dx 
\frac{1}{\cos^2(x)} \right] \label{solex}
\end{equation}
which is a solution of a differential equation
\begin{equation}
\frac{df}{dx} + \tan(x) f = \frac{1}{\cos(x)} \label{deex}
\end{equation}
The solution Eq. (\ref{solex}) is similar to Eq. (\ref{solq2}). It
also has a similar property with respect to initial values namely
$\cos(x_0)\neq 0$, where $x_0$ is the initial value, has to be
obeyed. Then
\begin{eqnarray}
f(x) = \frac{f(x_0) - \sin(x_0)}{\cos(x_0)} \, 
\cos(x) + \sin(x) = \cos(x) \left[ \frac{f(x_0)}{\cos(x_0)} 
+ \int_{x_0}^x dx \, \frac{1}{\cos^2(x)} \right] \nonumber \\
\label{solex2}
\end{eqnarray}
The first line shows that there are no problems for the solution at
any $x$ but the second form has the same problem as was seen in
Eq. (\ref{sol2q2}), namely there can be divergences within the range
of integration. Although, as long as we are able to determine the
integral function the divergences are of no consequence, but for
numerical evaluation they are a complication and have to be
regulated. Typically the regulation is done by taking a time average
of the oscillating function inside the integral. However, one should
be careful under what situations one should do this. For example
taking a time average in Eq. (\ref{solex2}), which would mean
replacing $\cos^2x\to 1/2$, would result into a linear growth of
$f(x)$, whereas the true solution is just oscillating. If the source
term on the other hand grows exponentially (multiply $\sin x$ with
$e^x$) there would be $e^x$ inside the integral, too. Now the time
average would result correctly into exponential growth of $f(x)$ and
the behavior of the amplitude can be traced even if the phase of
oscillation would remain unknown. A good check on whether one gets
wrong kind of results by time averaging is to use the solution at the
points when the oscillating function does vanish, since these can be
explicitly solved from the original equation. 

Multiply Eq. (\ref{deex}) with $\cos x$ and then take $\cos x = 0$. As
a result $f(x)=\pm 1$ at these points which is reproduced by the
solution, Eqs. (\ref{solex},\ref{solex2}), and is actually independent
of the initial conditions. The same can be applied to Eq. (\ref{q2})
after multiplying by $\varphi^{(0)\,'}$, which would correspond to the
original Einstein equation and then taking $\varphi^{(0)\,'}=0$. This
procedure would exactly give the behavior of $Q^{(2)}_{\varphi}$ at
these points. The behavior during time of these special points serves
as a check for the goodness of taking the time averaging.


\subsection{Super Hubble modes}\label{SHM}

%
%
%
%
%
%

The region of study we are interested in is the CMB scales. These
scales correspond to $\kappa^2 < 10^{-50}$
\cite{Bassett:1999cg}. Therefore for any practical purposes we can
take $\kappa^2=0$ as an approximation. This special case has the
benefit that analytical solution exists \cite{Ivanov}
\begin{equation}
\chi^{(1)}_{\sigma} = \left\{
\begin{array}{llll}
A^{(2n)} F(\alpha,\beta;\gamma: cn^4(y,1/\sqrt{2})) + 
B^{(2n)} cn(y,1/\sqrt{2}) \times \\
F(1+\alpha-\gamma,1+\beta-\gamma;2-\gamma: cn^4(y,1/\sqrt{2})), 
2n\omega \leq y \leq (2n+1)\omega \\
A^{(2n+1)} F(\alpha,\beta;\gamma: cn^4(y,1/\sqrt{2})) - 
B^{(2n+1)} cn(y,1/\sqrt{2}) \times \\
F(1+\alpha-\gamma,1+\beta-\gamma;2-\gamma: cn^4(y,1/\sqrt{2})), 
(2n+1)\omega \leq y \leq (2n+2)\omega
\end{array} \right. \label{chi1sol}
\end{equation}
where $y=x-x_0$, $F$ is the hypergeometric function and
\begin{equation}
\alpha = \frac{1}{8} \left(1+\sqrt{1+8 \frac{g^2}{\lambda}} \right), 
\quad \beta = \frac{1}{8} \left(1-\sqrt{1+8 \frac{g^2}{\lambda}} 
\right), \quad \gamma = 3/4, \quad n = 0, 1, 2, \ldots
\end{equation}
The coefficients $A^{(n)},B^{(n)}$ are related to each other by
\begin{eqnarray}
& \left( \begin{array}{ll} A^{(n+1)} \\ B^{(n+1)} 
\end{array} \right) = 
T \left( \begin{array}{ll} A^{(n)} \\ B^{(n)} \end{array} \right), 
\nonumber \\
& T = \left( \begin{array}{ll} \quad \sqrt{2} 
\cos\pi(\alpha-\beta) & -\frac{8\pi\Gamma(2-\gamma)^2}
{\Gamma(1-\alpha) \Gamma(1-\beta) \Gamma(1+\alpha-\gamma) 
\Gamma(1+\beta-\gamma)} \\ -\frac{8\pi\Gamma(\gamma)^2}
{\Gamma(\alpha) \Gamma(\beta) \Gamma(\gamma-\alpha) 
\Gamma(\gamma-\beta)} & \qquad\qquad \sqrt{2} \cos\pi(\alpha-\beta) 
\end{array} \right)
\end{eqnarray}
Using this recursion relation all the coefficients can be given in
terms of $A^{(0)},B^{(0})$. The eigenvalues of the matrix $T$ give the
characteristic exponents as a function of $s=g^2/\lambda$
\begin{equation}
\mu(s) = \frac{2\sqrt{\pi}}{\Gamma(1/4)^2} \, 
\ln\left\{ \sqrt{2} \left| \cos\left(\frac{\pi\sqrt{1+8s}}
{4}\right)\right| \pm \sqrt{\cos\left( \frac{\pi\sqrt{1+8s}}{2}\right)} 
\right\} \label{mus}
\end{equation}
The maximal value is obtained with values $s=2n^2-1/8$, where
$n=1,2,\ldots$, and is
\begin{equation}
\mu_{max} = \frac{2\sqrt{\pi}}{\Gamma(1/4)^2}\, \ln(1+\sqrt{2}) 
\approx 0.2377
\end{equation}
The initial coefficients $A^{(0)}, B^{(0)}$ are given in terms of
initial conditions as
\begin{eqnarray}
& \left( \begin{array}{ll} A^{(0)} \\ B^{(0)} \end{array} \right) 
= \tilde{T} \left( \begin{array}{ll} \chi^{(1)}_{\sigma}(x_0) \\ 
\frac{d \chi^{(1)}_{\sigma}}{d x}(x_0) \end{array} \right) \nonumber \\
& \tilde{T} = \pi^{3/2} \left( \begin{array}{ll} \frac{\sqrt{2}}
{\Gamma(\gamma)\Gamma(1+\alpha-\gamma)\Gamma(1+\beta-\gamma)} & 
\quad \frac{1}{2\Gamma(\gamma)\Gamma(1-\alpha)\Gamma(1-\beta)} \\ 
\quad -\frac{\sqrt{2}}{\Gamma(2-\gamma)\Gamma(\alpha)\Gamma(\beta)} 
& -\frac{1}{2\Gamma(2-\gamma)\Gamma(\gamma-\alpha)\Gamma(\gamma-\beta)}\
\end{array} \right) \label{tildet}
\end{eqnarray}
The initial values are obtained from the values they have at 
the end of inflation. Any massive field during inflation behaves as
\begin{equation}
\chi^{(1)}_{\sigma} = \frac{\pi}{2} \, e^{i(\nu_{\sigma}+1/2)\pi/2} \, \sqrt{-\eta} H^{(1)}_{\nu_{\sigma}} (-k\eta) = e^{i(\nu_{\sigma}-1/2)\pi/2} \, 2^{\nu_{\sigma}-3/2} \,\frac{\Gamma(\nu_{\sigma})}{\Gamma(3/2)} \, \frac{1}{\sqrt{2k}} \, (-k\eta)^{1/2-\nu_{\sigma}}
\end{equation}
where $\nu_{\sigma}=3/2 + \epsilon - \eta_{\sigma}$, where $\epsilon$
and $\eta_{\sigma}=m_{\sigma}/3H^2$ are slow-roll parameters,
$H^{(1)}_{\nu_{\sigma}}$ is a Hankel function and the second
expression applies to large scales. This results into
$\chi^{(1)\,'}_{\sigma} = {\cal H} (\nu_{\sigma}-1/2)
\chi^{(1)}_{\sigma}$ at large scales and therefore it also obeys the
condition for getting rid of the non-locality.


\end{document}